\documentstyle[epsf]{article}
\def\be{\begin{equation}}
\def\ee{\end{equation}}
\def\bea{\begin{eqnarray}}
\def\eea{\end{eqnarray}}
\def\LM{\Lambda}
\def\pa{\partial}

\def\ep{\epsilon}
\def\th{\theta}
\def\CF{{\cal F}}
\def\half{\frac{1}{2}}
\begin{document}
\begin{titlepage}
\setcounter{footnote}0
\begin{center}
\hfill Landau-96-TMP-3\\
\hfill ITEP-TH-17/96\\
\hfill hep-th/9607169\\
\vspace{0.3in}
\bigskip\bigskip
{\bf Akin $N=2$ SUSY Yang Mills
Theories and Instanton Expansion}\\
\bigskip\bigskip\bigskip
{Sergei Gukov \footnote{E-mail address: gukov@itp.ac.ru}
$^{\dag}$,
Igor Polyubin \footnote{E-mail address: polyub@itp.ac.ru}
$^{\ddag}$}\\
\bigskip
$\phantom{gh}^{\dag}${\em L.D.Landau Institute for
Theoretical Physics
\\ 2, Kosygin st., 117334, Moscow, Russia}\\
\bigskip
$\phantom{gh}^{\ddag}${\em L.D.Landau Institute for
Theoretical Physics
\\ 2, Kosygin st., 117334, Moscow, Russia\\
and\\
Institute of Theoretical and Experimental Physics\\
25, B.Cheremushkinskaya st., 117259, Moscow, Russia}\\
\bigskip
{\em 12 July 1996}\\ \end{center} \begin{abstract}

  The low energy effective actions of the $N=2$ SUSY
$SU(N_c)$ QCD are considered at the symmetric point on
the moduli space. The classes of such theories have similar
spectral curves.  This fact allows us to show that all
these models have the same structure of the coupling
matrix and to show that the $N_f=2N_c$ spectral curve can
not be presented as a double covering of the sphere. We
calculate first instanton contributions to the coupling matrix 
and get nonperturbative $\beta$-functions in the  $SU(2)$ 
gauge theory with non-zero bare masses of the matter 
hypermultiplets.

\end{abstract}

\end{titlepage}

\newpage

\setcounter{footnote}0

\section{Introduction}

\qquad Many exact results have been obtained
since the discovery by Seiberg and Witten \cite{sewi,sewi2}
of the exact solution of the low energy $N=2$
supersymmetric $SU(2)$ gauge theory in four dimensions.
The low energy $N=2$ SUSY Yang-Mills theories contain
$g=N_c-1$ abelian $N=2$ vector supermultiplets, which can
be decomposed into $g$ $N=1$ chiral multiplets $A_i$ plus
$g$ $N=1$ vector multiplets $W_i$.  According to
\cite{sewi,sewi2} the scalar component $a_i$ of the $A_i$
(namely, its  vacuum expectation value (vev)) is the local
coordinate on the quantum moduli space of the effective
action of the theory. It is given by the integrals
of meromorphic differentials
$\lambda$ over the basic cycles
$\alpha_{i}$ and $\beta_{i}$, such that $\alpha_{i} \circ
\beta_{j} = \delta_{ij}$, on a hyper-elliptic curve $\cal C$.
In particular, their derivatives with respect to the
symmetric vevs
$s_k=(-1)^k \sum_{i_1< \cdots < i_k} a_{i_1}
\cdots a_{i_k}$ ($k=2 ,
\cdots , N_c $) are equal to \cite{KlLeTh}:
\bea
{\pa a_i \over \pa s_k}
\sim A_{ki} = \int_{\alpha_{i}} \frac{x^{N_c-k}dx}{y}
\nonumber \\
{\pa a_Di
\over \pa s_k} \sim B_{ki} = \int_{\beta_{i}}
\frac{x^{N_c-k}dx}{y}
\label{ab}
\eea
\qquad We will be mainly interested in the coupling
matrix $T_{ij}$, which is the period matrix on ${\cal C}$.
In the matrix form, it can be presented as \cite{KlLeTh}:
\be
{\bf T} = {\bf A}^{-1}{\bf B} \label{tau}
\ee
\qquad In some particular cases the integration in
(\ref{ab}) may be performed explicitly. It has been done
for the $SU(2)$ \cite{KlLeTh,fipo,haoz} and $SU(3)$ groups
\cite{KlLeTh}. For higher $N_c$, it is not so immediate,
and, therefore, there were a series of calculations taking
into account only the one - instanton corrections
\cite{itsa}. For $N_c>3$, integration in (\ref{ab}) becomes
much more complicated, and Picard - Fuchs equations are not
known for arbitrary $N_c$ and $N_f$.

Some theories have the same
integrals - this allows us to derive relations between
couplings (see Section $2$) and, in some particular cases,
even to obtain exact results. The first step in this
direction was done in \cite{mine}, where some exact beta
functions for the $SU(2)$ and $SU(3)$ cases were calculated.

In Section 2, we discuss the coupling constants for the
theories with $N_c$ colors and $N_f$ ($N_f=0$ or 
$N_f=N_c$ ) massless flavours. We calculate them for 
the second order in the instanton expansion by the direct 
evaluation of integrals (\ref{ab}).

All the bare masses $m_k$ of the matter hypermultiplets
are put zero (till the Appendix).

As later we deal with instanton contributions to
the prepotential, it is relevant to remind the
perturbative expansion of the prepotential which is
saturated in one loop due to the supersymmetry:
\be
\CF=i \frac{2N_c-N_f}{8\pi}\sum_{i<j}(A_i-A_j)^2
\log{\frac{(A_i-A_j)^2}{\LM^2}}
\label{pprep}
\ee

In Section 3 we consider the scale invariant $N_f=2N_c$ 
theory. It has the classical period matrix ${\bf T}$ 
proportional to the matrix $\bf C$ : $C_{ij}=\delta_{ij}+1$ 
(in the basis $A_i=a_i$, $i=1 \cdots g$ ;
$A_{N_c}=-\sum_{i=1}^g a_i$). By comparing the spectral 
curves for this theory and for the $N_f=N_c$ one, we 
demonstrate that spectral curve for UV finite $N_f=2N_c$ 
($N_c>2$) theory can not be hyperelliptic 
(double covering of $CP^1$).

Section 4 is devoted to the $SU(2)$ theory. We present
the strong evidence for the method proposed by J. A. Minahan
and D. Nemeschansky which helps one to obtain some useful
relations between the $N_f=0$ and $N_f=2$ theories.

In Appendix we present some nonperturbative
$\beta$-functions of the $SU(2)$ gauge theory with non-zero
masses of the matter hypermultiplets.

\newpage

\section{General $N_c$}

\qquad Now we are going to discuss some akin $N=2$ SUSY
Yang-Mills theories which have the similar spectral curves
each depending on one dimensionless parameter. Namely, we
note that the curves for the $N_f=0$ and $N_f=N_c$ theories
have the same forms in the symmetric point on the moduli
space (compare (\ref{0nf}) and (\ref{nc})). The period
matrix $T_{ij}$ must be the same for the both theories,
turning into each other by an appropriate replace of the
parameters. Before going further, let us stress that the
curves are regarded akin if they are related by
$SL(2,{\bf C})$ transformation (this is the common
property of the two-dimensional manifolds) or by
rescaling $x$ and $y$ (because we restrict ourselves to the
only parameter and $T_{ij}$ is dimensionless too).

For instance, for general $N_c>2$ with all the order
parameters $s_k$ being zero but $s_{N_c}=-u \not= 0$, the
$N_f=0$ and $N_f=N_c$ curves take the forms \cite{haoz}:

\begin{itemize}
\item$N_f=0$
\be
y^2=\left(x^{N_c}-u^{(0)}\right)^2-\LM^{(0)2N_c}
\Leftrightarrow y^2=x^{2N_c}-2F^{(0)}x^{N_c}+1
\label{0nf}
\ee
\item$N_f=N_c$
\be
y^2=\left(x^{N_c}-u^{(N_c)}+\frac{\LM^{(N_c)N_c}}
{4}\right)^2-\LM^{(N_c)N_c}x^{N_c} \Leftrightarrow
y^2=x^{2N_c}-2F^{(N_c)}x^{N_c}+1
\label{nc}
\ee
\end{itemize}
where
\begin{itemize}
\item$N_f=0$
\be
F^{(0)}=\frac{u^{(0)}}{\sqrt{u^{(0)2}-\LM^{(0)2N_c}}}
\Leftrightarrow \LM^{(0)2N_c}=u^{(0)2}(1-F^{(0)-2})
\label{0fun}
\ee
\item$N_f=N_c$
\be
F^{(N_c)}=\frac{u+{\LM^{(N_c)N_c} \over 4}}
{u-{\LM^{(N_c)N_c} \over 4}}
\Leftrightarrow \LM^{(N_c)N_c}=4u^{(N_c)}
\frac{F^{(N_c)}-1}{F^{(N_c)}+1}
\label{ncfun}
\ee
\end{itemize}
As a consequence, their couplings are (here we denote
$v={\displaystyle\frac{\LM^{N_c}}{u}}$):
\bea
T_{ij}^{(N_c)}(v^{(N_c)})=T_{ij}^{(0)}\left(
\sqrt{\frac{v^{(N_c)}}
{(1+{v^{(N_c)} \over 4})^2 }}\right);\nonumber\\
T_{ij}^{(0)}(v^{(0)})=T_{ij}^{(N_c)}\left( \frac{8}{v^{(0)2}}
\left( 1 - \frac{v^{(0)2}}{2} - \sqrt{1 - v^{(0)2} }
\right)\right),
N_c>2
\label{vev}
\eea

The same procedure may be performed for other gauge groups.
One does not need to know integrals (\ref{ab}) explicitly.
Instead, it is sufficient to compare the curves, i. e. to
obtain relations between the quantities which may be
${\bf T}$, $\beta$, $\CF$ or their expansions in series at
small $v$.

Let us evaluate (\ref{ab}) to find period matrix, namely,
first instanton terms. We will focus on the $N_f=0$ theory
(as it  has been shown earlier $N_f=N_c$ case is completly
analogous):
\bea
A_{kl}=2\int_{x_{-,l}}^{x_{+,l}}\frac{x^{N_c-k}dx}{y}=
-\frac{2 \pi i}{N_c} \ep^{l(1-k)}z_{-}^{-m}F_{2,1}
(\half,m;1;1-\frac{z_{+}}{z_{-}}) \label{a} \\
B_{kl}=2\int_{x_{-,1}}^{x_{-,l}}\frac{x^{N_c-k}dx}{y}=
\frac{2}{N_c}(\ep^{l(1-k)}-\ep^{(1-k)})
\frac{z_{-}^{\half-m}}{z_{+}^{\half}}
\frac{\Gamma(1-m) \Gamma(\half)}{\Gamma({3 \over 2} - m)}
F_{2,1}(\half,1-m;{3 \over 2}-m;{z_{-} \over z_{+}})
\label{b}
\eea
\begin{figure}
\epsfxsize 400pt
\epsffile{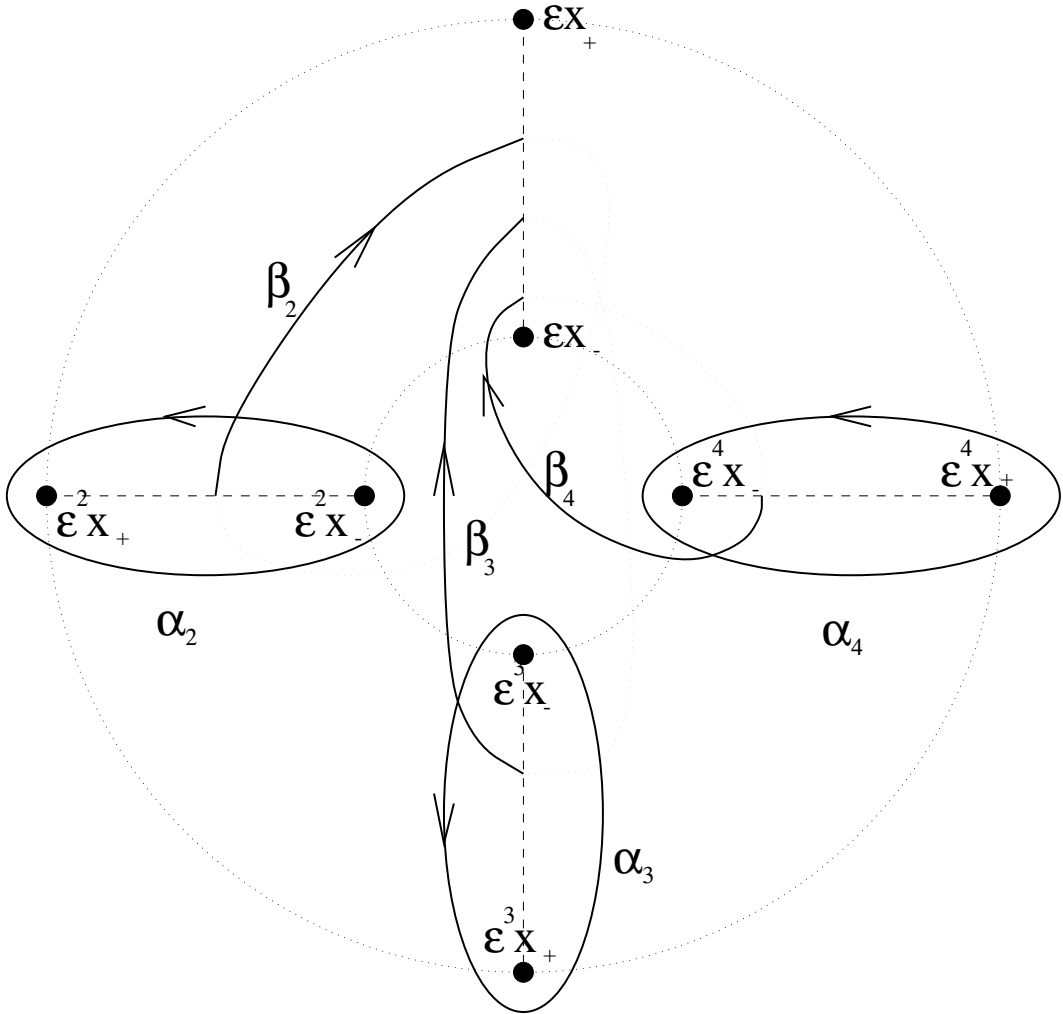}
\caption{ The homology cycles for $N_c=4$}
\end{figure}
where we chose the cycles as shown on Fig.1
($N_c=4$ assumed) so that:
\bea
x_{\pm,l}=\ep^l x_{\pm};
\ep=\exp\left( \frac{2 \pi i}{N_c}\right);m=\frac{k-1}{N_c}\\
x_{\pm}=(u \pm \LM^{N_c})^{{1 \over N_c}}=
z_{\pm}^{{1 \over N_c}};
\label{roots}\\
\lambda=\frac{z_{+}}{z_{-}}-1;
\eea
In the weak coupling limit ($ | \lambda | \ll 1$), 
(\ref{a}) and (\ref{b}) take the forms:
\be
A_{kl}=-\frac{2 \pi i}{N_c} \ep^{l(1-k)}z_{-}^{-m}(1-{m
\over 2} \lambda
+{3 \over 16}m(m+1)\lambda^2+ ({1 \over 8} - {3 \over 32}m
- {1 \over 6}m^2)m\lambda^3 +\cdots) \label{acl} \\
\ee
$$
B_{kl}=\frac{2}{N_c} (\ep^{l(1-k)}-\ep^{(1-k)})z_{-}^{-m}
( (\ln {4 \over \lambda} - C - \psi (1-m))(1 - {m\over 2}
\lambda +{3 \over 16}m(m+1)\lambda^2  -
$$
\be
- {5 \over 96}m(m+1)(m+2)\lambda^3 ) + {\lambda \over 2} +
{m^2-5m - 3 \over 16} \lambda^2 +
({5 \over 48}+{1 \over 32}m(8+2m-m^2))\lambda^3
+ \cdots )
\label{bcl}
\ee
where $C$ is the Euler constant, and $\psi(x)=
{\Gamma' \over \Gamma}$ is
the logarithmic derivative of the gamma function.
It is useful to note that the dependence of ${\bf A}$
and ${\bf B}$ on $k$ and $l$ is :
\bea
A_{kl}=a(k) E_{kl} \qquad E_{kl}=\ep^{l(1-k)}
\label{AA}
\\
B_{kl}=b(k)(\ep^{l(1-k)}-\ep^{(1-k)})=b(k)(E_{kl}+
\sum_{l}E_{kl})
\label{AB}
\eea
Substituting it into (\ref{tau}), one gets the matrix
of coupling constants for the $N_f=0$ theory:
\be
T_{ij}=\sum_{k}{b(k) \over a(k)}(E^{-1})_{ik}(E_{kj}+
\sum_{j}E_{kj}) = 
\label{TT}
\ee
$$
={i \over \pi}\sum_{k}(E^{-1})_{ik}(E_{kj}
+\sum_{j}E_{kj})( \ln {4 \over \lambda}
-C -\psi (1-m) + {\lambda \over 2} -
{3 \over 16}\lambda^2 +\cdots )
$$

It is easy to show that ${\bf T}$ is proportional
to ${\bf C}$, if and only if all the exact
$a(k)^{-1}b(k)$ are the same functions for any $\it k$.
Imposing the constraint $T_{ii}=2T_{i \not= j}$ on
${\bf T}$, one immediately gets from (\ref{TT}):
$$
\sum_{k} \frac{b(k)}{a(k)} (E^{-1})_{ik} E_{kj} |_{i
\not= j}= 0
$$
which, after simple transformations, turns into :
\be
a^{-1}(k)b(k)=const
\label{test}
\ee
This equation must be satisfied in all orders in $\lambda$.
Obviously, it is not true and ${\bf T} \not\sim {\bf C}$
in symmetric point.

\section{$N_f=2N_c$ case}

\qquad When $N_f=2N_c$ and the bare masses are zero we 
get conformally invariant theory. It has the classical period
matrix ${\bf T}$ proportional to the matrix ${\bf C}$ :
$T_{ij}=\tau C_{ij}=\tau (\delta_{ij} + 1)$.

The spectral curve for this case was proposed in
\cite{haoz} and, when the masses are set to zero,
curve reads ($s_i=0$, $i \not= N_c$; $s_{N_c}=-u \not= 0$):
\footnote{$L$ and $l$ are modular forms expressed
through the higher genus $\th$-constants defined on
$\tau {\bf C}$}
\be
y^2=\left[(1+{L\over 4})x^{N_c}-ul\right]^2-Lx^{2N_c}
\Leftrightarrow y^2=x^{2N_c}-2Fx^{N_c}+1
\label{2nc}
\ee

This curve has the same form as (\ref{0nf})
and (\ref{nc}), with the associated period matrix 
${\bf T} \sim {\bf C}$ having the same structure. From the 
computation of Section 2 one can easily see that this is not 
the case. Moreover, it may be seen from (\ref{pprep}) that 
even the perturbative coupling matrix is not proportional to 
the matrix ${\bf C}$ anywhere on the moduli space if 
$N_c>2$ and $N_f<2N_c$ (see also \cite{mine}).

The simplest
way to see it is to compute ${\bf T}$ in basis
$a_i=A_i-A_{N_c}$ , $i=1 \cdots g$. Requirement for $T_{ij}$
to be proportional to the classical
matrix leads to the constraint on $a_i$ :
$$
(g-1) \log {a_i} = \sum_{k \not= i}^g \log (a_i-a_k)
$$
which must be satisfied for any $i$. Since these equations
have not nontrivial solutions,
we come to the statement that ${\bf T} \not \sim {\bf C}$
at any point on the moduli space.

Furthemore, let us suppose that the
$N_f=2N_c$ curve is written as a polynomial of power
$2N_c$ (for instance, as it was proposed in \cite{haoz}).
One can compare the spectral curve for such a theory and 
for the $N_f=N_c$ one. Since the both theories have the 
spectral curves which are polynomials of power $2N_c$ and 
there are $3N_c-2$ parameters ($s_k^{(N_c)}$, $s_k^{(2N_c)}$
and $m_k^{(N_c)}$), one can adjust them so that the
curves are getting identical (up to $SL(2,C)$ transformations), 
and the corresponding theories have the same structure of the 
coupling matrices. But from the previous arguments based on 
perturbative results, we know that it does not take place. 
Hence, the $N_f=2N_c$ spectral curve is not a polynomial of 
power $2N_c$.

Thus, we demonstrate that, for $N_f=2N_c$, the spectral curve
can not be hyperelliptic surface (the double covering of
$CP^1$). Our conjecture is that, for scale invariant
theories, covering of the sphere must be replaced by
covering of elliptic curve with natural elliptic parameter
$\tau = {\th \over 2\pi} + {4 \pi i \over g^2}$.

\section{More on $N_c=2$}

\qquad Let us repeat the same procedure for the $SU(2)$
group in detail. Analogously:
\bea
T^{(N_c)}(v^{(N_c)})=T^{(0)}\left(\sqrt{\frac{v^{(N_c)}
(1+\frac{v^{(N_c)}}{8}) }{(1+{3v^{(N_c)} \over 8})^2 }}
\right);
\nonumber\\
T^{(0)}(v^{(0)})=T^{(N_c)}\left( 8 \frac{1 - \sqrt{1 -
v^{(0)2}}} {1 + 3 \sqrt{1 - v^{(0)2}}} \right),N_c=2
\label{rel}
\eea
We use it to relate the coefficients $\CF_k$ of the
instanton expansion for the prepotential:
\be
\CF (a)=\frac{ia^2}{4 \pi} \left[ b \ln \left({a \over \LM}
\right) + \sum_{k=1}^{\infty} \CF_k (N_f)\left({\LM \over a}
\right)^{kb} \right]
\label{prep}
\ee
where $b=4-N_f$. The order parameter is known from
the Picard-Fuchs equation \cite{Ito}:
\be
u=\frac{4 \pi}{ib}\left( a{\pa \CF \over \pa a} -2\CF\right)=
a^2 \left[ 1-\sum_{k=1}^{\infty} k\CF_k
\left({\LM \over a}\right)^{kb} \right]
\label{order}
\ee
$T$ appears as the second derivative of the prepotential :
\be
T = {i \over 2 \pi} \left[ b \left( {3 \over 2}+
\ln \left({a \over \LM}\right)\right)+\sum_{k=1}^{\infty}
(1- {kb \over 2})
(1 - kb) \CF_k \left( \frac{\LM}{a} \right) ^{kb}\right]
\label{tau2}
\ee

Substituting $N_f=0$ and $N_f=2 (=N_c)$ and inverting the
series for $u$, one gets $a$:
\begin{itemize}
\item $N_{f}=0$
\be
\frac{a^2}{\LM^2}=\frac{u}{\LM^2} \left[ 1+
\CF_1 \left( {\LM^2 \over u} \right)^2 +(2\CF_2- \CF_1^2)
\left( {\LM^2 \over u} \right) ^4 + (3 \CF_3 - 8 \CF_1 \CF_2 +
2 \CF_1^3) \left( { \LM^2 \over u} \right)^6 + \cdots \right]
\ee
\item $N_f=2$
\be
\frac{a^2}{\LM^2}=\frac{u}{\LM^2}\left[ 1+
\CF_1 \left( {\LM^2 \over u}\right) +2\CF_2 \left( {\LM^2
\over u}\right)^2 + (3\CF_3-2\CF_1\CF_2)\left(
{\LM^2 \over u}\right)^3 +\cdots \right]
\ee
\end{itemize}
After insertion of these results into (\ref{tau2}), we
obtain the instanton expansion for $T$
($v={\displaystyle{\LM^2 \over u}}$):
\begin{itemize}
\item $N_{f}=0$
\bea
T^{(0)} =\frac{i}{2\pi} \left[ 6+ \ln {a^2 \over \LM^2} +
3\CF_1^{(0)} \left( {\LM^2 \over a^2} \right)^2 +21\CF_2^{(0)}
\left( {\LM^2 \over a^2} \right)^4+55\CF_3^{(0)}
\left( {\LM^2 \over a^2}
\right)^6 + \cdots \right] = \nonumber \\  =
\frac{i}{2\pi} \left[ 6-
\ln v^{(0)2} +4\CF_1^{(0)}v^{(0)2}+(23\CF_2^{(0)}-
{15 \over 2}\CF_1^{(0)2})v^{(0)4}+ \cdots \right]
\label{tau20}
\eea
\item $N_f=2$
\bea
T^{(2)} = \frac{i}{2\pi} \left[ 3+ \ln {a^2 \over \LM^2} +
3\CF_2^{(2)} \left( {\LM^2 \over a^2} \right)^2 +
10\CF_3^{(2)}\left( {\LM^2 \over a^2} \right)^3 +
\cdots \right] = \nonumber
\\ = \frac{i}{2\pi} \left[ 3- \ln v^{(2)} +\CF_1^{(2)}v^{(2)}+
(5\CF_2^{(2)}+ \half \CF_1^{(2)2})v^{(2)2}+ \cdots \right]
\label{tau22}
\eea
\end{itemize}

In order to check (\ref{rel}), one must substitute
$$
8 \frac{1 - \sqrt{1 - v^{(0)2}}}{1 + 3 \sqrt{1 - v^{(0)2}}} =
v^{(0)2}\left( 1+{5 \over 8} v^{(0)2} + {29 \over 64}v^{(0)4}
+\cdots \right)
$$
into (\ref{tau22}):
\bea
T^{(0)}(v^{(0)})=T^{(N_c)}\left( 8 \frac{1 -
\sqrt{1 - v^{(0)2}}}{1 + 3 \sqrt{1 - v^{(0)2}}} \right)
= \nonumber\\ \frac{i}{2\pi} \left[ 3- \ln v^{(0)2} +
(\CF_1^{(2)}-{5 \over 8}) v^{(0)2}+(5\CF_2^{(2)}+
{5 \over 8} \CF_1^{(2)} + \half \CF_1^{(2)2} -
{1 \over 16})v^{(0)4} + \cdots \right]
\eea
Comparison with the coefficients in (\ref{tau20})
yields the system of equations for the first two of them:
\begin{eqnarray}
\left\{
\begin{array}{ccc}
4\CF_1^{(0)}=\CF_1^{(2)}-{5 \over 8}\nonumber\\
23\CF_2^{(0)}-{15 \over 2} \CF_1^{(0)2}=5\CF_2^{(2)}+
{5 \over 8}\CF_1^{(2)}+\half \CF_1^{(2)2}-{1 \over 16}
\end{array}
\right.
\label{system}
\end{eqnarray}
\qquad It may seem not to be so interesting to deal with
these equations, since one can get explicit expressions
for the coefficients in terms of $\th$-constants for
these theories, nevertheless it allows to express
immediately $N_f=2$ instanton terms through $N_f=0$ ones
and provides the good consistency check for them. Let us
mention that the results of \cite{Ito} do not satisfy them.

Now let us prove the exact formula expressing
$\beta$ through $\th$ - constants \footnote{$\th_2=
\th[\half,0]=\sum_{n \in Z}q^{(n+ \half)^2}$,
$\th_3=\th[0,0]=\sum_{n \in Z}q^{n^2}$, $\th_4=
\th[0,\half]=\sum_{n \in Z} (-1)^n q^{n^2}$} \cite{mine}:
\bea
\beta^{(0)} = {2 \over \pi i} \frac{\th_3^4(2T) +
\th_2^4(2T)}{\th_4^8(2T)} \nonumber \\
\beta^{(2)} = {1 \over 2 \pi i} \frac{\th_3^4(2T) +
\th_4^4(2T)}{\th_3^4(2T) \th_4^4(2T)}
\label{su2}
\eea
The $N_f =4$ curve is \cite{haoz}:
\bea
y^2=x^4-2Fx^2+1 \nonumber\\
F=\frac{\th_2^4(T) + \th_3^4(T)}{\th_2^4(T) -
\th_3^4(T)}
\label{nf4}
\eea
where $x$ and $y$ are rescaled by the vev
$u$. Of course, $a$, $a_D$, $A$
and $B$ depend on $u$ but $T$.
So we know the connection between the factor $F$ in equation
(\ref{nf4}) for $u$ and $T$. Let us show it by direct
calculation for the $N_f=0$ curve \cite{haoz}:
\be
y^2=(x^2-u)^2-\LM^4
\label{nf0}
\ee
This equation may be rewritten in the form :
\be
y^2=x^4-2 \frac{u}{\sqrt{u^2-\LM^4}} x^2+1
\label{nf01}
\ee
{}from which (using
(\ref{nf4})) we get $\LM^4=u^2(1-F^{-2})$ (the result of
\cite{mine}), or, finally:
\be
\LM^4=u^2(1-F^{-2})=u^2 \left({2\theta_2^2 \theta_3^2 \over
\theta_2^4+\theta_3^4}\right)^2
\label{sol0}
\ee

One can easily integrate (\ref{ab}) in
terms of hyper-geometric functions:
\bea
A = 2 \int_{x_{+}}^{x_{-}} \frac{dx}{\sqrt{(x^2-u)^2-\LM^4}}
= - {2\pi i \over x_{+}+x_{-}}F_{2,1}(\half,\half;1;{(x_{+}-
x_{-})^2 \over (x_{+}+x_{-})^2})\nonumber\\
B = 2 \int_{-x_{+}}^{x_{+}} \frac{dx}{\sqrt{(x^2-u)^2-\LM^4}}
= {2\pi \over x_{-}}F_{2,1}(\half,\half;1;\left({x_{+}
\over x_{-}}\right)^2)
\label{per}
\eea
where $x_{+}=\sqrt{u+\LM^2}$ and $x_{-}=\sqrt{u-\LM^2}$
are roots of (\ref{nf0}). Note also that $B$ has the form:
\be
B = {2\pi \over x_{-}}F_{2,1}(\half,\half;1;\left({x_{+}
\over x_{-}} \right)^2) = {2\pi \over x_{+}+x_{-}}F_{2,1}
(\half,\half;1;{4 x_{+} x_{-}\over (x_{+} + x_{-})^2})
\label{per1}
\ee
As it was mentioned above, $T$ is the ratio of periods:
$$
T =i { F_{2,1}(\half,\half;1;1-w)
\over F_{2,1}(\half,\half;1;w) }
$$
with $w ={\displaystyle \frac{(x_{+}-x_{-})^2}{(x_{+}+x_{-})^2}}$.
Solution
to this equation is known (see \cite{baer} for example):
$$ w =
\frac{\th_2^4(0,q)}{\th_3^4(0,q)} \Rightarrow u-\LM^2 = (u+\LM^2)
\left(\frac{\th_3^2-\th_2^2}{\th_3^2+\th_2^2}\right)^2
$$
which leads to the
result (\ref{sol0}) after simple transformations.  We have
obtained it by comparing this $N_f=0$ curve (\ref{nf0}) and
that with $N_f=4$ (\ref{nf4}) (the method by Minahan and
Nemechansky \cite{mine}) and by direct calculation, providing
the evidence for the validity of this method (by comparing
the curves). In this way, we can get the exact
$\beta$ - functions (see also Appendix A).  They are in
agreement with \cite{KlLeTh,fipo}, but not with \cite{Ma}.

\section{Conclusions}
\qquad We find first instanton corrections to the  matrix
of coupling constants at the symmetric point on the moduli
space. Comparing the spectral curves for the theories with
different number of flavors, we get some useful relations
between couplings and present the proof that the $N_f=2N_c$
spectral curve could not be presented as a covering of
the sphere. Also we propose the strong evidence for the
method of \cite{mine} (comparing the curves) in the case
of the $SU(2)$ gauge group. One can easily extend this
technique to the theories with nonzero bare masses, but
the requirement for the coupling matrix to be proportional
to the classical one imposes some constraints on the masses
(or on their symmetric functions $t_k(m)=\sum_{i_1 < \cdots
< i_k} m_{i_1} \cdots m_{i_k}$). Some $\beta$ - functions
for such theories are collected in the Appendix. It is easy
to check that the results turns into (\ref{su2}) if the
masses tend to zero.

\vskip5mm
\section*{Acknowledgments}

\qquad We are indebted to A. Morozov for
helpful discussions and comments.
S. Gukov would like to thank A. Mironov and S. Khoroshkin
for teaching him some useful mathematical background.
I. Polyubin is grateful to the Institute of Theoretical
Physics at Hannover for kind hospitality where part of
this work was done.

The work of S.G. was partially supported by RFBR
grant No. 95-01-00755, the work of I.P. by grants
RFBR-96-01-01106 and INTAS-93-1038. I. Polyubin also
would like to thank Volkswagen Stiftung project "Integrable
models and strings" for financial support.

\section*{Appendix A : Some exact $\beta$ - functions at
nonzero bare masses}
\begin{itemize}
\item $N_c=N_f=2$, $m_1=-m_2=m$
$$
\LM^2=(F^2-9)^{-1}\left[ 8F^2(u-4m^2)+24u-8
\sqrt{8F^4m^2(2m^2-u)+8uF^2(2u-3m^2)} \right]
$$
so that
$$
\beta= \frac{F^2-9}{FF'} \left[ \frac{ \left[ u-4m^2-
\frac{4F^2m^2(2m^2-u)+2u(2u-3m^2)}{ \sqrt{2F^4m^2(2m^2-u)+
2uF^2(2u-3m^2)}} \right](F^2-9)}{F^2(u-4m^2)+3u-
\sqrt{8F^4m^2(2m^2-u)+8uF^2(2u-3m^2)}} - 1 \right]^{-1}
$$
where for the $SU(2)$ group :
$$
F=\frac{\th_2^4(T) + \th_3^4(T)}{\th_2^4(T) -
\th_3^4(T)}
$$

\item $N_c=2$, $N_f=3$, $m_1^2+m_2^2=2u$, $m_3=0$
$$
\beta=\frac{F+2}{F'} - \frac{4(F+2)^2 \left( u +
\frac{\LM^2}{64} \right)^2}{27F'(u-m_1^2)^2}
$$
$$
\frac{F+2}{27}\left( u + \frac{\LM^2}{64} \right)^3=
\frac{\LM^2}{64}(u-m_1^2)^2
$$
where
$$
F=\frac{(2\th_4^4+\th_2^4)(2\th_2^4+\th_4^4)(\th_4^4-
\th_2^4)}{(\th_2^8 + \th_4^8 + \th_2^4\th_4^4)^{3 \over 2}}
$$
\end{itemize}

\end{document}